\documentclass[10pt,twocolumn,letterpaper]{article}

\usepackage{cvpr}              

\newcommand{\ours}{\textsc{Ours}\xspace}
\newcommand{\pidef}{\textsc{PI-DEF}\xspace}

\newcommand{\bhnerf}{{BH-NeRF}\xspace}
\newcommand{\mlp}{{4D-MLP}\xspace}

\usepackage{amsmath,amsfonts,bm}
\usepackage{dsfont}

\def\eqref#1{equation~\ref{#1}}

\def\1{\bm{1}}

\def\vtheta{{\bm{\theta}}}

\def\vb{{\bm{b}}}

\DeclareMathAlphabet{\mathsfit}{\encodingdefault}{\sfdefault}{m}{sl}
\SetMathAlphabet{\mathsfit}{bold}{\encodingdefault}{\sfdefault}{bx}{n}

\definecolor{cvprblue}{rgb}{0.21,0.49,0.74}
\usepackage[pagebackref,breaklinks,colorlinks,allcolors=cvprblue]{hyperref}
\usepackage{titletoc}

\def\paperID{38441} 
\def\confName{CVPR}
\def\confYear{2026}

\title{Dynamic Black-hole Emission Tomography with Physics-informed Neural Fields}

\author{
Berthy T.~Feng$^{1,2,3}$ \quad Andrew A.~Chael$^{4,5}$ \quad David Bromley$^{6}$ \\
Aviad Levis$^{6}$ \quad William T.~Freeman$^{2,3}$  \quad Katherine L.~Bouman$^{1}$\\
$^{1}$Caltech \quad $^{2}$MIT \quad $^{3}$NSF IAIFI \quad $^{4}$ Princeton University \\ $^{5}$ Niels Bohr International Academy \quad $^{6}$University of Toronto
}

\begin{document}
\maketitle
\begin{abstract}
With the success of static black-hole imaging, the next frontier is the dynamic and 3D imaging of black holes.
Recovering the dynamic 3D gas near a black hole would reveal previously-unseen parts of the universe and inform new physics models.
However, only sparse radio measurements from a single viewpoint are possible, making the dynamic 3D reconstruction problem significantly ill-posed.
Previously, BH-NeRF addressed the ill-posed problem by assuming Keplerian dynamics of the gas, but this assumption breaks down near the black hole, where the strong gravitational pull of the black hole and increased electromagnetic activity complicate fluid dynamics.
To overcome the restrictive assumptions of BH-NeRF, we propose \textit{PI-DEF}, a physics-informed approach that uses differentiable neural rendering to fit a 4D (time + 3D) emissivity field given EHT measurements.
Our approach jointly reconstructs the 3D velocity field with the 4D emissivity field and enforces the velocity as a soft constraint on the dynamics of the emissivity.
In experiments on simulated data, we find significantly improved reconstruction accuracy over both BH-NeRF and a physics-agnostic approach.
We demonstrate how our method may be used to estimate other physics parameters of the black hole, such as its spin.
\end{abstract}    
\section{Introduction}
\label{sec:intro}
\begin{figure}
    \centering
    \includegraphics[width=\linewidth]{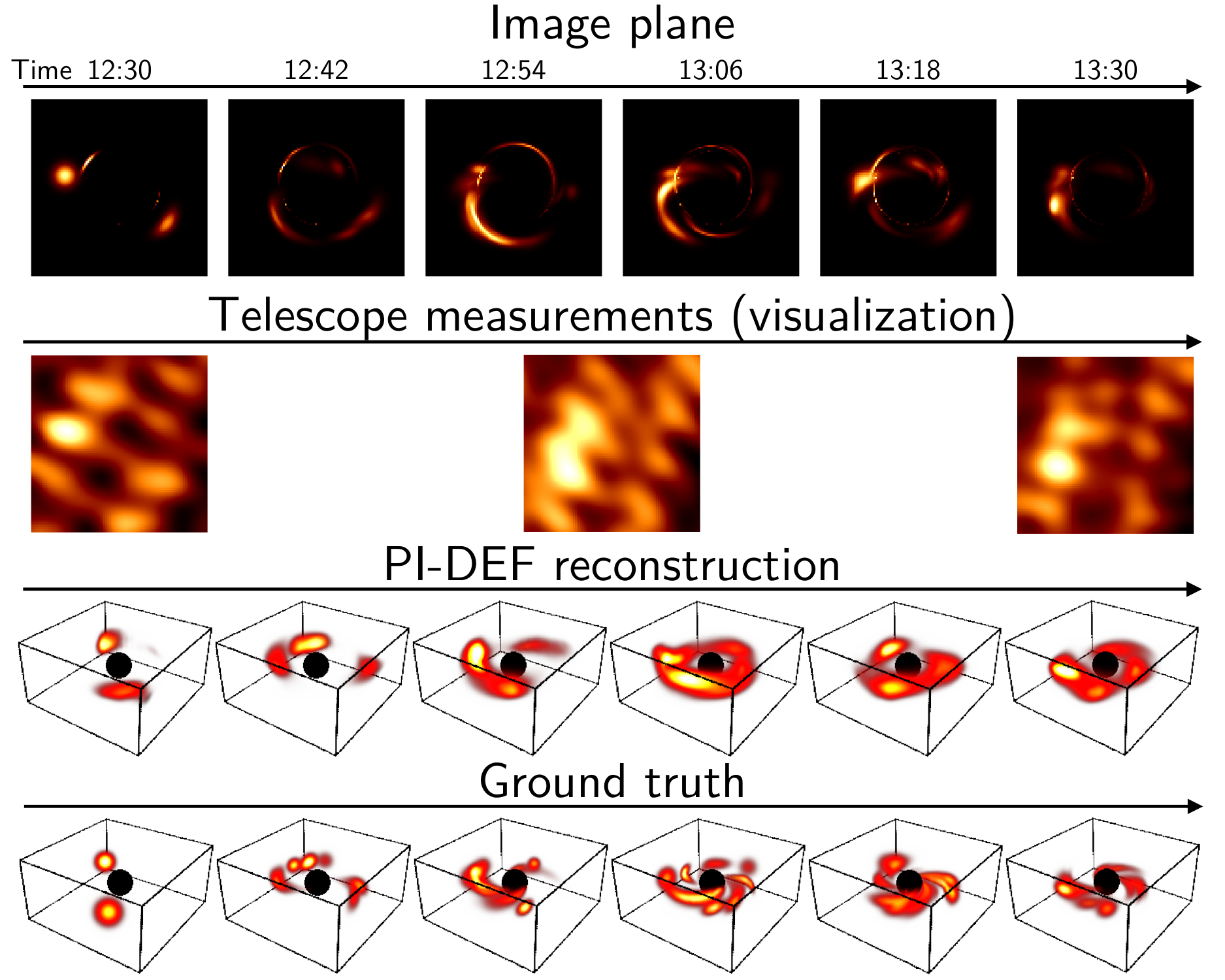}
    \caption{Our approach solves a severely ill-posed tomography problem in astrophysics. Given sparse EHT telescope measurements (visualized here as images), we recover a 4D emissivity field of the moving gas near a black hole. Our approach imposes soft physics constraints to constrain the reconstruction and recovers a 3D velocity field that explains the observed motion.
    \vspace{-20pt}
    }
    \label{fig:teaser}
\end{figure}
Through the imaging of the black holes M87* and Sagittarius A* (Sgr A*), the Event Horizon Telescope (EHT) demonstrated that it is possible to image distant black holes by combining data from radio telescopes distributed across Earth.
The resulting images \citep{m87paperiv,sgrapaperiii} of both black holes corroborated a key prediction of Einstein’s theory of relativity: the presence of a persistent, circular black-hole ``shadow.''
However, static 2D images provide an incomplete view of the dynamic 3D environment around the black hole, limiting our ability to test the theory more deeply.
An image is a complicated 2D projection of the 3D emissivity around the black hole and thus may obscure important 3D structures.
Moreover, a static image does not capture the dynamic nature of the radiation-emitting gas, which moves, appears, and disappears over time.
Visualizing the dynamic 3D emissivity field near a black hole would push the boundaries of scientific knowledge by revealing previously-unseen parts of the universe and informing physics models.

Dynamic 3D black-hole imaging is challenging for several reasons.
First, the EHT can only observe the black hole from a single viewpoint, and its measurements are highly sparse and corrupted, meaning that the inverse problem is highly ill-posed.
Second, the source is evolving, meaning that EHT measurements cannot be simply aggregated across time to improve reconstruction quality.
Third, the way light gets propagated to the image plane depends partially on the unknown fluid dynamics near the black hole, meaning that the measurement forward model in our inverse problem is not fully known.
With these challenges, we require a method that incorporates assumptions that are informative enough to constrain the solution space, yet relaxed enough that they can help us overcome unknown physics.

Prior to our work, the only existing method for black-hole emission tomography was BH-NeRF \citep{levis2022gravitationally}, which models the initial 3D emissivity as a neural field and propagates it according to Keplerian dynamics.
BH-NeRF poses two key limitations.
First, since the neural field only models the initial emissivity, BH-NeRF cannot account for new emission over the time of imaging.
Second, BH-NeRF will struggle to recover emissivity fields that do not respect Keplerian velocities.
A technique based on BH-NeRF was used to perform the first 3D recovery of the gas around a black hole \citep{levis2024orbital}, specifically a flare near Sgr A*.
However, it was only effective in that setting because the flare was sufficiently far away from the black hole so that its dynamics could be described as Keplerian, and it was suspected that no new flares appeared during the observation time used for imaging.
In contrast, we target emission much closer to the black hole, where hotspots of emission appear and disappear much more rapidly, and the fluid velocity fluctuates much more from the Keplerian model.

We propose \textit{\pidef} (physics-informed dynamic emission fields), a physics-informed approach that simultaneously reconstructs the dynamic 3D emissivity field and the 3D velocity field of the emitting gas while imposing soft physics constraints on both fields.
Specifically, we represent both fields as neural fields via coordinate-based neural networks and optimize them according to physics-informed losses.
By reconstructing a 4D (i.e., time-dependent 3D) field, we are able to account for new emissions.
By enforcing physics as a soft constraint, we are robust to differences between the assumed velocity field and the real-world velocity field, which may include turbulence, non-zero radial velocities, and sub-Keplerian speeds.
Our approach offers the dual benefits of achieving a more accurate dynamic 3D emissivity reconstruction \textit{and} inferring the unknown velocity field near a black hole.

In this paper, we first discuss related work in tomographic imaging and computational methods and then cover relevant physics background.
We then present \pidef and results showing its superior performance in terms of both emissivity and velocity recovery compared to BH-NeRF and a physics-agnostic method.
We showcase the possibility of inferring the spin of the black hole using our method.

Our work highlights the crucial role of computer vision in solving fundamental physics problems.
Computer vision helped us attain the first picture of a black hole \citep{bouman2016computational} and the first 3D recovery of a flare near a black hole \citep{levis2024orbital}.
We apply computer vision techniques -- specifically in computational imaging and graphics -- to see even further near black holes.
Eventually, visual reconstructions from real data may have the potential to help scientists test fundamental theories of general relativity and quantum mechanics.
The continued involvement of the computer vision community \citep{eht_tutorial} is therefore essential for answering the most challenging and important questions in physics.
\section{Related work}
\label{sec:related_workd}
\subsection{Tomography}
Tomography is a type of inverse problem that aims to recover an object or structure from its lower-dimensional projections.
A common tomographic problem is reconstructing a 3D image from 2D projections at multiple viewpoints, as in computed tomography \citep{kak_principles_2001} for medical imaging and 3D scene reconstruction from photos \citep{mildenhall_nerf_2020}.
These settings, however, benefit from linear ray-tracing, multiple viewpoints, and the assumption of a static source.
In our case, we work with curved light paths, a single viewpoint, and a dynamic source.
Essentially, we aim to solve a 4D tomography problem by recovering a time-dependent 3D image.

Previous efforts in many scientific domains have dealt with problems related to non-linear light paths.
For example, imaging underwater environments \citep{xiong_in-the-wild_2021}, imaging through the Earth's atmosphere \citep{levis_multiple-scattering_2017}, and imaging cosmological objects \citep{hoekstra_weak_2008} are difficult due to refraction, scattering, and gravitational lensing, respectively.
We deal with curved light paths due to strong gravitational lensing, an effect of curved spacetime around a black hole.
In the context of astronomical imaging, deprojecting galaxies \citep{rybicki_deprojection_1987,zhao_single_2024,zhao_revealing_2024} from 2D images is a common problem, but the galaxies are usually parameterized as static, simple shapes.
Previous work recovered hotspots of emission around a black hole using EHT data \citep{tiede_spacetime_2020,levis2024orbital}, but they assumed a simplified geometric model \citep{tiede_spacetime_2020} or a simplified dynamics model \citep{levis2022gravitationally} for the hotspots.
In contrast, we use a time-dependent neural field to capture complicated features and dynamics.

\subsection{Coordinate-based neural fields}
Instead of representing a volume with a discretized representation, we can represent it with a coordinate-based neural network.
Coordinate-based neural fields are parameter-efficient, continuous representations that use a multilayer perceptron (MLP) to map from the coordinates of a point to the value of the field at that point.
They also benefit from the implicit regularization of a neural network that leads to smoothness in the output field.
They are commonly used to represent solutions to inverse problems, where the weights of the MLP are optimized to minimize a data-fit loss.
For example, a neural radiance field (NeRF) \citep{mildenhall_nerf_2020} is trained to represent a 3D scene that agrees with a sparse set of 2D views, thus solving a 2D-to-3D tomography problem with linear ray-tracing.
Similar approaches have been developed for scientific imaging tasks, such as cryo-electron microscopy \citep{zhong_cryodrgn_2021}, MRI \citep{tancik_fourier_2020}, and computed tomography \citep{sun_coil_2021}.

Coordinate-based neural fields have been applied to dynamic scenes.
A straightforward approach is to add a time dimension to the input coordinates \citep{gao_monocular_2022,xian_space-time_2021,li_neural_2021,gao_dynamic_2021}.
Another approach is to use one neural representation for the scene at the first time frame and a second neural representation for the time-dependent deformation \citep{pumarola_d-nerf_2021,tretschk_non-rigid_2021,park_nerfies_2021,park_hypernerf_2021}.
Our work combines both approaches by estimating both a time-dependent emissivity field and a velocity field describing its dynamics.
When measurements are sparse, the main challenge is finding a temporal prior to describe how points in space change across time.
Previous methods regularized the deformation to be sparse or rigid \citep{park_nerfies_2021,tretschk_non-rigid_2021}, while others imposed models highly specialized to the object being reconstructed \citep{levis2022gravitationally,li_neural_2021,peng_animatable_2021,gafni_dynamic_2021}.
We propose a way to impose a soft physics-based constraint on the velocity field.

We note that Gaussian splatting \citep{kerbl_3d_2023,luiten_dynamic_2024} has emerged as an alternative efficient and continuous representation.
This approach represents a 3D density distribution as a set of Gaussians with learnable positions, scales, and orientations.
Its main benefit is that rendering can be made extremely efficient.
However, the number of Gaussians needs to be pre-defined, so it is unsuitable for our setting, in which hotspots of emission may appear and disappear over time.

\subsection{BH-NeRF}
Our work builds upon BH-NeRF \citep{levis2022gravitationally}, which represents the initial 3D emissivity with a coordinate-based neural field, denoted by $e_0(\mathbf{x})$.
The weights of $e_0$ are optimized to minimize a data-fit loss with respect to time-dependent EHT measurements.
For a given time $t$, the optimization algorithm propagates $e_0$ according to Keplerian orbital velocities to that time, passes it through the EHT forward model, and compares the estimated measurements with the observed measurements at time $t$.


BH-NeRF encounters two major limitations: (1) it cannot account for emission that appears after the start of observation, and (2) it cannot account for velocity that is different from the assumed velocity.
For (1), because BH-NeRF only estimates the initial 3D emissivity, it cannot account for new emission appearing in the volume over time.
For (2), the Keplerian model does not account for infall into the black hole or other considerations such as turbulence.
Velocity becomes less and less Keplerian as we get closer to the black hole, where high-energy effects and geodesic effects causing gas to fall into the black hole become more prevalent.
Furthermore, imposing the assumed velocity model as a strong constraint makes it difficult to overcome modeling errors.
Our work addresses these limitations, recovering a time-dependent emissivity field and a velocity field.
We impose velocity as a soft constraint, meaning our method is robust to errors in the assumed velocity model.
Since our emissivity field is time-dependent, it can capture new emission across the entire observation time window.
\section{Background}
\label{sec:background}
\subsection{Black-hole emission physics}
\label{sec:blackhole_physics}
Emission near a black hole can come from either inflowing gas in an accretion disk or outflowing gas from a jet \citep{abramowicz_foundations_2013}.
The gas radiates, emitting photons that travel through space.
We assume the photons follow a certain fluid velocity and are ray-traced along geodesics to determine the emission observed on the image plane.
The supplementary text contains more details about the relevant physics.

\subsubsection{Fluid velocity}
\label{sec:fluid_velocity}
We denote the spherical velocity vector as $v^i$:
\begin{equation}
\label{eq:vi}
    v^i=\left(v^r, v^\theta, v^\phi\right) = \left(\frac{\mathrm{d}r}{\mathrm{d}t},\frac{\mathrm{d}\theta}{\mathrm{d}t},\frac{\mathrm{d}\phi}{\mathrm{d}t}\right),
\end{equation}
where $i$ indexes into the spherical coordinates $(r,\theta,\phi)$, which denote the radius, polar angle, and azimuthal angle, respectively.
In general relativity, we often work with the four-velocity vector $u^\mu$, where $\mu\in\{0,1,2,3\}$ is a spacetime index. Conventionally, $0$ and $(1,2,3)$ correspond to time and spatial coordinates, respectively.
The four-velocity represents velocity relative to the proper time $\tau$:
\begin{equation}
\label{eq:umu}
u^\mu = \left(\frac{\mathrm{d}t}{\mathrm{d}\tau}, \frac{\mathrm{d}r}{\mathrm{d}\tau},\frac{\mathrm{d}\theta}{\mathrm{d}\tau},\frac{\mathrm{d}\phi}{\mathrm{d}\tau}\right) = u^t\left(1,v^r,v^\theta,v^\phi\right),
\end{equation}
since $u^t=\mathrm{d}t/\mathrm{d}\tau$ by definition.
The parameter $t$ is the coordinate time, or the time that a stationary observer at radius $r\to\infty$ measures.
The parameter $\tau$ is the proper time, which is what the particle moving with velocity $u^\mu$ measures.
The four-velocity $u^\mu$ can be directly derived from $v^i$ by computing $u^t$ to satisfy the normalization condition $u^\mu u_\mu =-1$.
Since there are only three degrees of freedom, our method only needs to estimate the coordinate three-velocity $v^i$.

\paragraph{General fluid velocity model}
Keplerian motion is a simple velocity model that assumes a constant orbital radius.
Close to the black hole, however, matter plunges inward along geodesics, and motion is not exactly Keplerian.
For a more general velocity model, we turn to the velocity model proposed by \citet{cardenas-avendano_adaptive_2023} for adaptive analytical ray-tracing (AART), which accounts for sub-Keplerian velocity and infall.
Given a fixed spin $a$, mass $M$, and direction of orbit (prograde or retrograde) of the black hole, the AART velocity model involves just three parameters: $\beta_\phi$, $\beta_r$, and $\xi$.
The parameters $\beta_\phi,\beta_r\in[0,1]$ represent the relative magnitude of the sub-Keplerian velocity in the angular and radial velocities, respectively, where $\beta_\phi=\beta_r=1$ corresponds to a purely sub-Keplerian model, and $\beta_\phi=\beta_r=0$ corresponds to pure infall.
The sub-Keplerianity parameter $\xi\in(0,1]$ represents the ratio of the angular momentum to the Keplerian value.

\subsubsection{Normal observer frame}
\label{sec:normal_observer}
We represent the velocity field as a coordinate-based neural network, which may estimate unreasonable velocities that would cause the equation for $u^t$ to be undefined.
To avoid numerical errors, we instead estimate velocities in the more numerically-stable \textit{normal observer} frame.
In this frame, we denote the three-velocity as $\tilde{u}^i$.
The supplementary text provides the formulae for converting between $\tilde{u}^i$ and $u^\mu$.

\subsubsection{Image formation}
\label{sec:image_formation}
The light captured on the image plane depends on two main ingredients: the emissivity field and the velocity field.
The emissivity $e$ determines how much intensity is integrated along light paths.
We drop its dependence on the frequency and direction \citep{levis2022gravitationally} and denote the emissivity field as a function of time and space: $e(t,\mathbf{x})$, where $\mathbf{x}=(x,y,z)$.

\vspace{-10pt}
\paragraph{Redshift}
Due to the Doppler effect, the frequency of light observed on the image plane is different from that of the original emission.
The velocity field dictates how the light intensity is boosted (redshift) or dimmed (blueshift) as it travels to the observer.
A source moving towards us near the speed of light appears brighter, whereas a source moving away appears dimmer.
The redshift factor $g$ directly depends on the four-velocity $u^\mu$ at each spatial location $\mathbf{x}$.

\vspace{-10pt}
\paragraph{Geodesic ray-tracing}
Radiative transfer describes how light propagates through a medium.
To make an image, we have to integrate a general relativistic radiative transfer (GRRT) equation along the curved light paths, known as \textit{geodesics}.
We make two simplifying assumptions in the GRRT equation that have been made in previous work \citep{levis2022gravitationally}.
First, we assume that the attenuation of light due to absorption and scattering is negligible, as is the case for EHT images \citep{m87paperv}. 
Second, we neglect the finite light-travel time, during which the relativistically-moving gas would shift position \citep{chan_gray_2013,dexter_public_2016,moscibrodzka_ipole_2018}.
With these assumptions, we can simply integrate the emissivity along the ray paths that end at the image plane.
Supposing a discretized image plane with $N\times N$ pixels, let $\Gamma_n=(t(s),\mathbf{x}(s))$ denote the geodesic that ends at the $n$-th pixel, where $s$ is the distance along the geodesic.
The observed intensity at the $n$-th pixel at time $t$ can be computed as
\begin{align}
\label{eq:raytracing}
    p_n(t)=\int_{\Gamma_n} g^2(\mathbf{x})e(t,\mathbf{x}) \mathrm{d}s.
\end{align}

\subsection{EHT measurements}
The EHT comprises multiple radio telescopes across Earth, which together collect measurements of the sky's image through very-long-baseline interferometry (VLBI).
Each pair of telescopes $i,j$, known as a \textit{baseline}, provides a Fourier measurement of the image, known as a \textit{visibility} $v_{ij}$ \citep{van1934wahrscheinliche,zernike1938concept,thompson2017interferometry}.
EHT measurements are challenging to invert into an image because they only sparsely sample the 2D Fourier plane.
In this work, we focus on fitting these complex visibilities (\cref{subsec:realistic} covers a more challenging type of measurement known as a closure phase).
Conditioned on an image $\mathbf{x}$, the measurement distribution can be modeled as Gaussian with the log likelihood
\begin{align}
    \log p(\mathbf{y}\mid\mathbf{x}) = -\frac{1}{2\mathbf{\sigma}^2}\left\lVert\mathbf{A}\mathbf{x}-\mathbf{y}\right\rVert_2^2,
\end{align}
where $\mathbf{A}$ denotes the linear forward model that sparsely samples Fourier measurements of the image $\mathbf{x}$, and $\mathbf{y}$ and $\sigma$ denote the measured visibilities and the standard deviations of their Gaussian noise, respectively.

\section{Method}
\label{sec:method}
\subsection{Representing the emissivity and velocity fields}
\begin{figure}
    \centering
    \includegraphics[width=\linewidth]{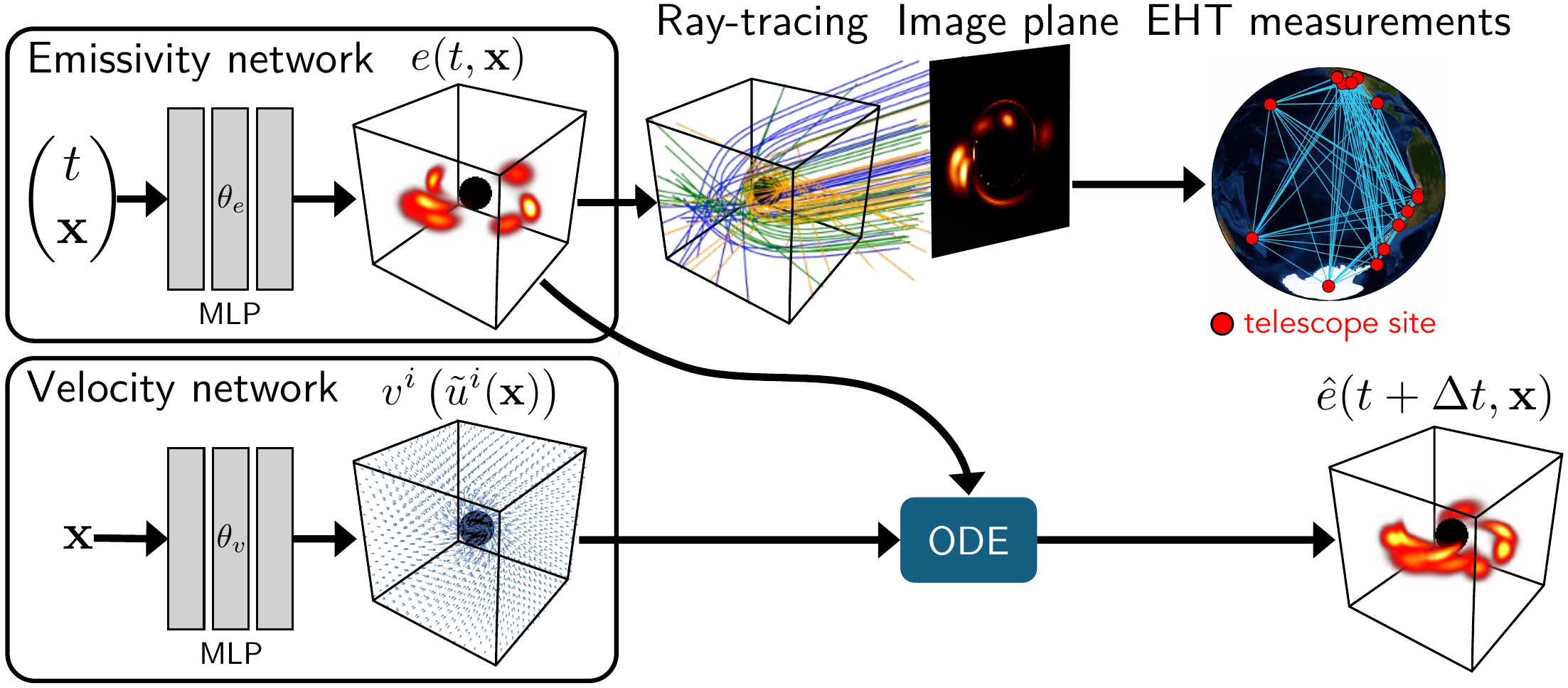}
    \caption[Black-hole emission tomography method overview.]{Main components overview. For every time $t$, we project the estimated emissivity $e(t,\mathbf{x})$ onto the image plane and simulate EHT measurements. A data-fit loss makes sure that the resulting measurements agree with the observed EHT measurements. We propagate $e(t,\mathbf{x})$ forward in time by $\Delta t$ via the velocity network and an ODE solver. A dynamics loss checks that the resulting $\hat{e}(t+\Delta t,\mathbf{x})$ agrees with the $e(t+\Delta t,\mathbf{x})$ given by the emissivity network. A velocity regularization loss supervises the velocity network with an assumed velocity model.}
    \label{fig:method}
\end{figure}

We denote the estimated emissivity and velocity by $e(t,\mathbf{x})$ and $\tilde{u}^i(\mathbf{x})$, respectively, where
\begin{align}
    e(t,\mathbf{x}) = e(t,\mathbf{x};\vtheta_e)&=\text{MLP}\left(\gamma\left([t,\mathbf{x}]^\top\right);\vtheta_e\right), \\
    \tilde{u}^i(\mathbf{x}) = \tilde{u}^i(\mathbf{x};\vtheta_v) &= \text{MLP}\left(\gamma(\mathbf{x});\vtheta_v\right).
\end{align}
Here $\vtheta_e$ and $\vtheta_v$ denote the parameters of the coordinate-based neural networks.
The positional encoding $\gamma$ has been shown to improve the representation of high-dimensional features \citep{tancik_fourier_2020} and is defined as
\begin{align}
    \gamma(\mathbf{x})=\left[\sin(\mathbf{x}),\cos(\mathbf{x}),\ldots,\sin\left(2^{L-1}\mathbf{x}\right),\cos\left(2^{L-1}\mathbf{x}\right)\right]^\top,
\end{align}
where the degree $L$ determines the bandwidth of the interpolation kernel \citep{jacot_neural_2018} underlying the MLP (i.e., higher $L$ allows for higher-frequency representations).
Note that $\tilde{u}^i$ is the three-vector in the normal observer frame defined in \cref{sec:normal_observer}.
\cref{fig:method} illustrates the optimization procedure for both networks, which we detail in the following subsection.

\subsection{Optimization}

We write the complete loss function as follows:
\begin{multline}
    \mathcal{L}\left(e,\tilde{u}^i\right)=\lambda_\mathrm{data}\mathcal{L}_\mathrm{data}\left(e,\tilde{u}^i\right)\\+\lambda_\mathrm{dyn}\mathcal{L}_\mathrm{dyn}\left(e,\tilde{u}^i\right)+\lambda_\mathrm{reg}\mathcal{L}_\mathrm{reg}\left(\tilde{u}^i\right).
\end{multline}

The \textbf{data-fit loss} encourages the emissivity network to fit the observed EHT measurements.
The forward model depends on the estimated velocity through the redshift factor, so this loss term depends on both networks:
\begin{equation}
\label{eq:data_fit_loss}
    \mathcal{L}_\mathrm{data}\left(e,\tilde{u}^i\right)=\sum_{n=1}^{N_t}\frac{1}{2\mathbf{\sigma}_n^2}\left\lVert \mathbf{A}_n\mathbf{x}\left(e(t_n);\tilde{u}^i\right)-\mathbf{y}_n\right\rVert_2^2,
\end{equation}
where $N_t$ is the number of time frames, $\mathbf{y}_n$ is the set of measured visibilities at time $t_n$, $\mathbf{A}_n$ is the corresponding EHT forward model, $\sigma_n$ is the corresponding set of Gaussian noise standard deviations, and $\mathbf{x}$ is the result of ray-tracing the 3D emissivity $e(t_n)$ onto the image plane with the redshift factor derived from the estimated velocity $\tilde{u}^i$.

The \textbf{dynamics loss} connects the emissivity and velocity networks by imposing a soft velocity constraint on the emissivity dynamics.
We define it as 
\begin{multline}
    \label{eq:dynamics_loss}
    \mathcal{L}_\mathrm{dyn}\left(e,\tilde{u}^i\right) \\= \mathbb{E}_{t\sim\mathcal{U}([0,T])}\left\lVert (\mathcal{G}\ast e)(t+\Delta t)-(\mathcal{G}\ast\hat{e})(t+\Delta t) \right\rVert_1,
\end{multline}
where $\Delta t$ is a small time interval, and $T$ is the total observation time.
The propagated emissivity as predicted by the velocity network is denoted by $\hat{e}(t+\Delta t)$, and it is given by
\begin{align}
    \hat{e}(t+\Delta t) &= e(t, \mathbf{x}-\hat{\mathbf{x}}(\Delta t)),
\end{align}
where
\begin{equation}
\label{eq:integral}
    \hat{\mathbf{x}} = \int_0^{\Delta t} v^i\left(\tilde{u}^i(\mathbf{x}(t)\right) \mathrm{d}t.
\end{equation}
In words, $\hat{e}(t+\Delta t)$ propagates the predicted emissivity at time $t$ forward in time by $\Delta t$, using the prediction of the velocity network.
\cref{eq:dynamics_loss} compares this to the actual emissivity field at time $t+\Delta t$, encouraging the time evolution in the emissivity network to agree with the velocity network.
In \cref{eq:dynamics_loss} we convolve $e$ and $\hat{e}$ with a Gaussian kernel $\mathcal{G}$ with standard deviation $\sigma=1.5$ to blur the estimated emissivities before evaluating the loss.
We find that blurring the emissivity and using an L1 loss are both helpful in preventing blurry reconstructions that cheat the dynamics loss.
 
We incorporate an assumed theoretical velocity model via \textbf{velocity regularization}:
\begin{align}
    \mathcal{L}_\mathrm{reg}(\tilde{u}^i) = \left\lVert v^i\left(\tilde{u}^i\right)-v^i_\mathrm{prior} \right\rVert_2^2,
\end{align}
where we compute $v^i_\mathrm{prior}$ from the AART velocity model detailed in \cref{sec:fluid_velocity}, and $v^i\left(\tilde{u}^i\right)$ is the three-velocity derived from the output of the velocity network in the normal observer frame.
As we discuss in \cref{sec:results}, we lower the regularization strength throughout optimization.
Regularization helps guide the networks at the beginning, but it becomes negligible by the end of optimization.
At that point, the networks are primarily fit to the measurements.

\subsection{Simulating ground-truth data}
We model hotspots in the same way as \citet{levis2022gravitationally}, where a hotspot is either a Gaussian blob with a certain position and scale or a tube with a certain position, length, and scale.
To simulate an emissivity field, we define a total number of flare events $N_\mathrm{flare}$, which occur at evenly spaced time intervals throughout the simulation, which lasts for a total time $T$.
Each flare event consists of $1$ to $3$ (sampled uniformly) individual hotspots.
For each hotspot, we randomly sample a radius $r\sim\mathcal{U}([r_{\min},r_{\max}])$ and scale $\sigma\sim\mathcal{U}([\sigma_{\min},\sigma_{\max}])$.
The hotspot has probability $1/2$ of being either a Gaussian blob or a tube.
If it is a Gaussian blob, then we randomly sample an azimuthal position $\phi\sim\mathcal{U}([0,2\pi])$.
If it is a tube, then we randomly sample an arc length $\ell\sim\mathcal{U}([0,\pi/5])$ and starting azimuthal position $\phi_0\sim\mathcal{U}([0,2\pi-\ell])$.
We solve an ODE involving the injected hotspots and the AART velocity model to determine the emissivity field at every time $t$.
To form the video, we perform geodesic ray-tracing at every time frame according to \cref{eq:raytracing} with the AART redshift factor.

\section{Results}
\label{sec:results}
Our results in simulation verify the superior reconstruction performance of \pidef compared to \bhnerf and a physics-agnostic baseline.
We verify that we recover the correct velocity in regions with enough moving gas, even when assuming an incorrect velocity model.
Looking forward to applying \pidef to real EHT data, we consider realistic Gaussian noise and atmospheric noise, and we show the potential of using \pidef to estimate important physics parameters such as the spin of the black hole.

\begin{figure*}
    \centering
    \includegraphics[width=0.85\linewidth]{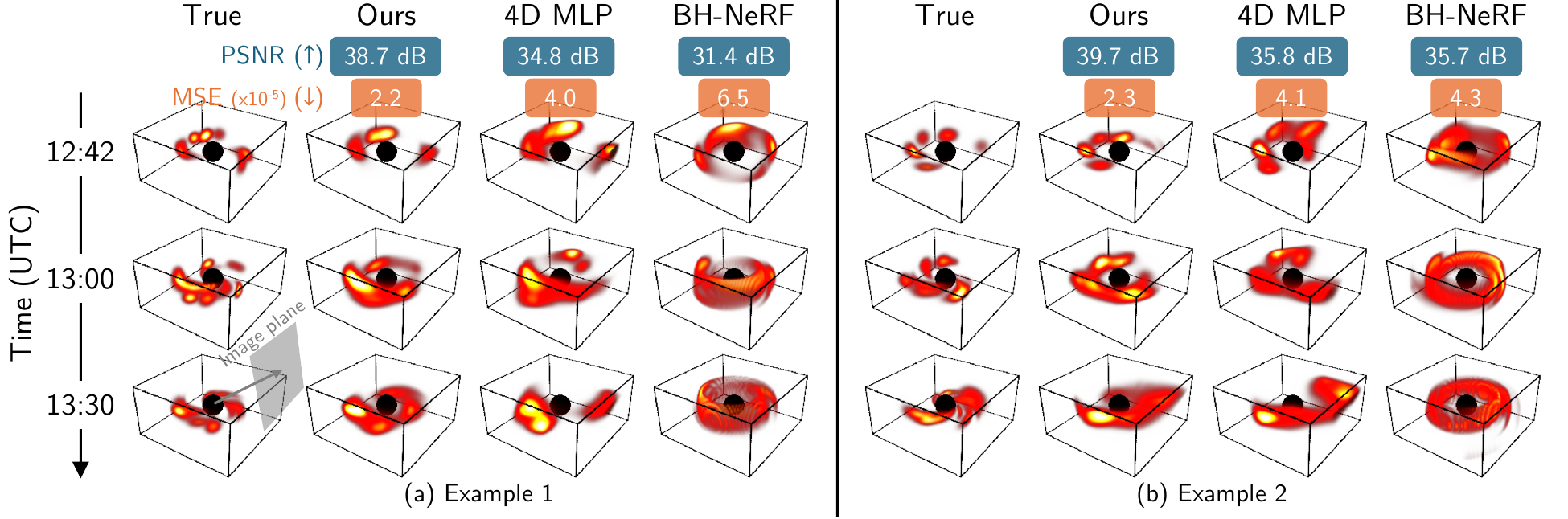}
    \caption{Example reconstructions of two random simulated ground-truth emissivity fields. \mlp is a physics-agnostic approach (i.e., the 4D MLP is just fit to the data). \bhnerf \citep{levis2022gravitationally} is the previous approach that enforces a strict Keplerian velocity prior, which is incorrect in this region near the black hole. Our physics-informed approach gives the most accurate reconstructions.}
    \label{fig:emissivity}
\end{figure*}

\subsection{Implementation}
\label{sec:implementation}
\cref{app:implementation} provides more details about network architecture, optimization, and data simulation.

\vspace{-10pt}
\paragraph{Optimization}
We used the Adam optimizer with the learning rate linearly decaying from $10^{-4}$ to $10^{-6}$ over $100$K iterations and a batch size of $6$.
We exponentially decayed the regularization weight according to
\begin{equation}
    \lambda(s)=\lambda_\mathrm{final}+(\lambda_{\mathrm{init}}-\lambda_{\mathrm{final}})\cdot\exp(-d\cdot s),
\end{equation}
where $s$ is the iteration, and we set the decay rate to $d=0.001$.
We set $\lambda_\mathrm{data}=1$ and $\Delta t =0.01$ for the dynamics loss term.
When fitting to EHT measurements, we set $\lambda_\mathrm{init}=10^6$ and $\lambda_\mathrm{final}=10$, and we set $\lambda_\mathrm{dyn}=10^5$.
When fitting to full images, we set $\lambda_\mathrm{init}=1$ and $\lambda_\mathrm{final}=10^{-6}$, and we set $\lambda_\mathrm{dyn}=0.1$.
We approximate the data-fit loss and velocity regularization by evaluating the neural fields on the geodesic points computed by \texttt{kgeo} \citep{kgeo}.
We evaluate the dynamics loss on an evenly-spaced $64\times 64\times 64$ grid ranging from $-12.5$ M to $12.5$ M in each direction.

\vspace{-10pt}
\paragraph{Simulating data}
We assumed a spin of $a=0.2$, inclination angle of $\theta_\mathrm{o}=60^\circ$, and FOV of $25$ M.
We randomly sampled hotspots and propagated them according to the AART velocity model with $\xi=0.7$ and $\beta=\beta_r=\beta_\phi=0.9$.
We ray-traced the emissivity field onto a $100\times 100$ image plane using the geodesics computed with \texttt{kgeo} \citep{kgeo}.
We used the \texttt{eht-imaging} library to simulate EHT observations taken during a one-hour observation window from $12.5$ to $13.5$ hours in UTC.
We assumed $102$-second-long scans, each $102$ seconds apart, where each scan provides a set of measured visibilities.
Our simulated emissivity fields have enough total flux, or brightness, to overcome the Gaussian noise of the measurements.
In \cref{subsec:realistic}, we consider measurements with realistic Gaussian noise.

\subsection{Baselines}
We compare to \textbf{\bhnerf} \citep{levis2022gravitationally} (the previous method for our task) and \textbf{\mlp} (a physics-agnostic baseline).
With \mlp, we optimize the emissivity network only to minimize the data loss (assuming a redshift factor based on the true velocity).
For our method, we used a mismatched velocity prior by assuming that $\xi=1$ and $\beta=1$.
This corresponds to a fully sub-Keplerian velocity model without any radial infall, which is close to the Keplerian assumption of \bhnerf.
As the following results show, unlike \bhnerf, our method can overcome this flawed velocity assumption.

\subsection{Emissivity accuracy}
\cref{fig:emissivity} shows example emissivity reconstructions of simulated ground-truth emissivity fields, comparing the different methods when using next-generation EHT (ngEHT) measurements.
We calculated PSNR by normalizing both the estimated and true emissivity 4D data volume to the range $[0,1]$.
Our method, even when assuming a mismatched velocity prior, provides significantly more accurate emissivity reconstructions than \mlp and \bhnerf.
\cref{tab:emissivity_metrics} verifies this trend in terms of PSNR and MSE evaluated on a test dataset of five randomly-generated emissivity fields.


\begin{table}
\centering
\caption{Emissivity reconstruction accuracy. Metrics are computed on the 4D emissivity volume. We report the mean $\pm$ std.~dev.~across a test dataset of five emissivity fields.}
{\small
\begin{tabular}{lcccc}
\toprule
 & PSNR (dB) & MSE ($\times 10^{-5}$) \\
\midrule
\ours & $\mathbf{37.3}\pm 2.3$ & $\mathbf{2.3}\pm 0.2$ \\
\mlp & $35.4\pm 0.5$ & $3.8\pm 0.4$ \\
\bhnerf & $34.0\pm 1.9$ & $4.9\pm 0.8$ \\
\bottomrule
\end{tabular}
}
\label{tab:emissivity_metrics}
\end{table}

\subsection{Measurement sparsity}
\begin{figure}
    \centering
    \includegraphics[width=\linewidth]{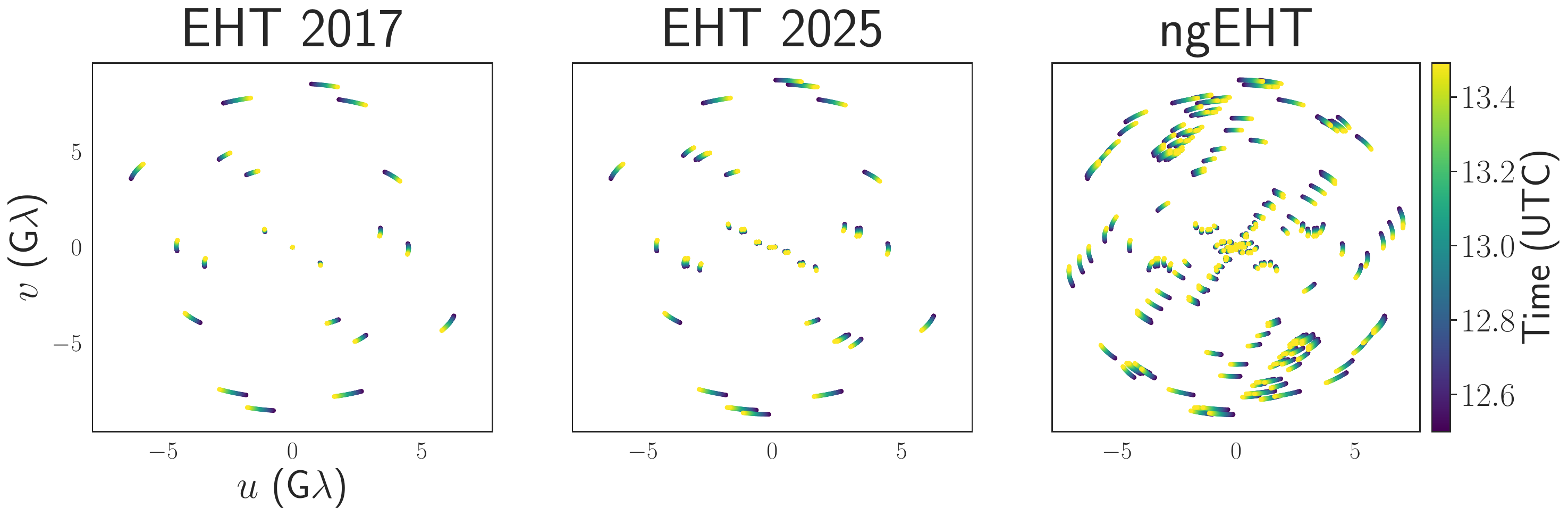}
    \caption{The $(u,v)$-coverage of the three arrays considered in our experiments. EHT 2025 marginally improves upon EHT 2017, while ngEHT significantly improves coverage. When simulating measurements, we kept the same bandwidth (2 GHz) for all three arrays, although ngEHT should have greater bandwidth (16 GHz).}
    \label{fig:uv_coverage}
\end{figure}

The EHT improves its measurement resolution by adding telescope sites.
The 2017 array used to capture the first image of M87* comprised eight telescopes, whereas the 2025 array has 12 telescopes,\footnote{We simulated measurements based on the proposed 2025 array that contains OVRO, but as of the time of writing, OVRO is still in the process of being commissioned.} and the ngEHT is slated to include 23 telescopes.
\cref{fig:uv_coverage} compares the $(u,v)$-coverage of the 2017, 2025, and ngEHT arrays.

\begin{figure}
    \centering
    \includegraphics[width=\linewidth]{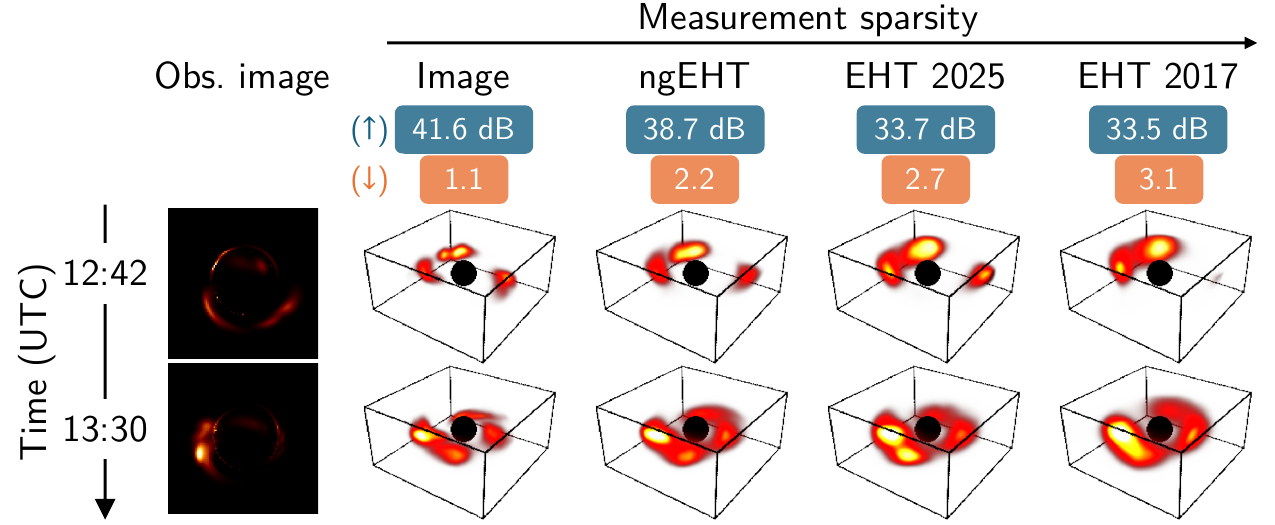}
    \caption{Emissivity reconstructions as measurements become sparser. ``Image'' corresponds to directly fitting to ``Obs.~image.'' 
    \vspace{-15pt}}
    \label{fig:meas_sparsity}
\end{figure}

\cref{fig:meas_sparsity} shows emissivity reconstructions of the same ground truth as in \cref{fig:emissivity}(a) but assuming different types of measurements: full images, ngEHT, EHT 2025, and EHT 2017.
We find that EHT 2025 does not improve upon EHT 2017 by much, while ngEHT provides a significant improvement in the recovered emissivity, achieving a resolution closer to that of full images, which represent the highest-possible measurement resolution from Earth.

\subsection{Velocity accuracy}
\begin{figure}
    \centering
    \includegraphics[width=\linewidth]{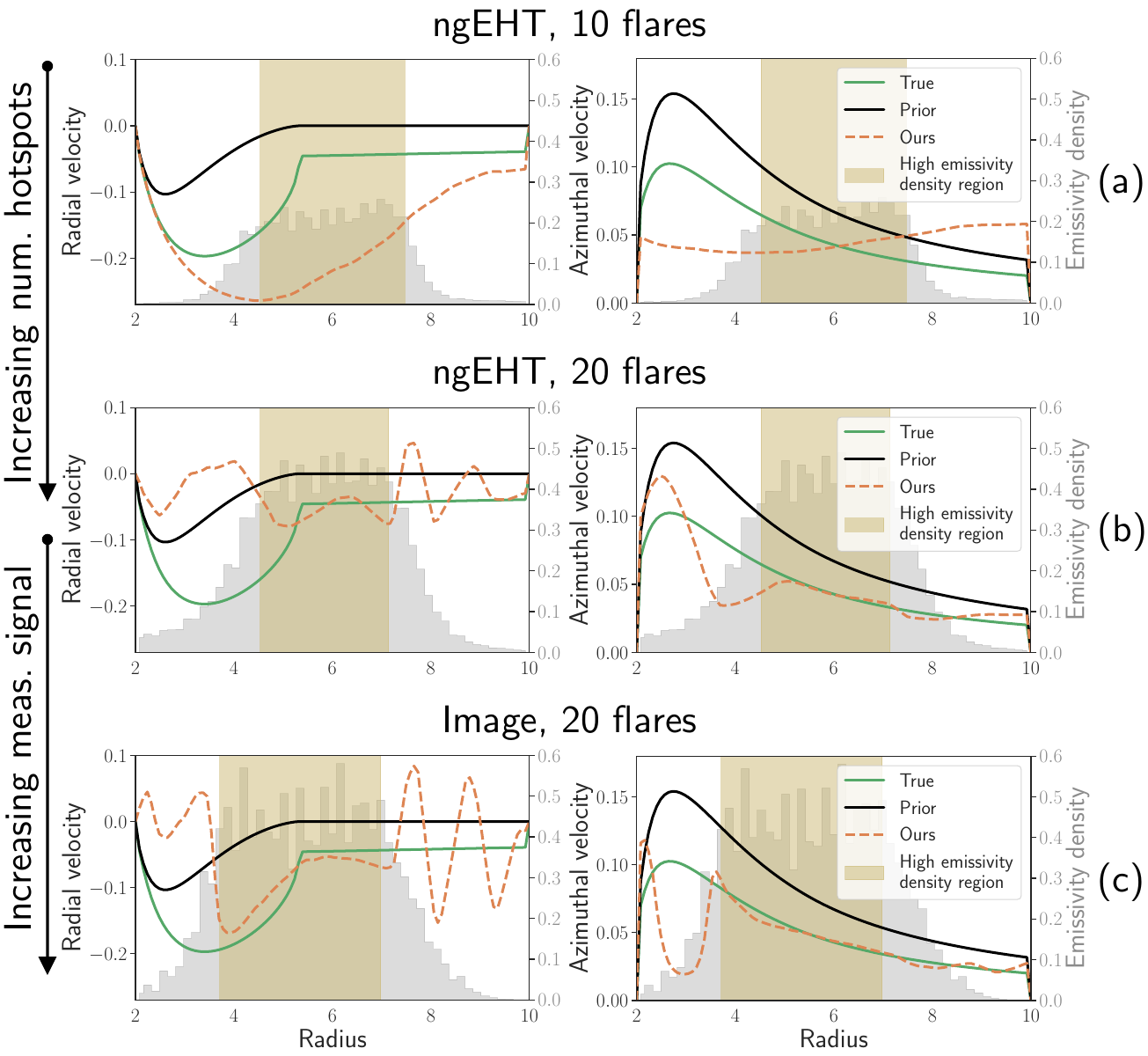}
    \caption{Velocity reconstruction depends on amount of recoverable emissivity. This could come from more emissions (e.g., going from 10 flares to 20 flares) or better measurements (e.g., going from ngEHT to image measurements). The gray histogram shows the density of the reconstructed emissivity (averaged across $z\in[-2,2]$ M and $\phi\in[0,2\pi]$ for every radius $r\in[0,10]$ M). The gold region indicates the region of high emissivity, defined as being above the $65$th percentile of all emissivity densities when grouped into $50$ evenly-spaced bins between $r=0$ and $r=10$. The recovered velocity is most accurate where there is most emissivity.
    Note how increasing the number of hotspots and the measured signal both increase the histogram height.
    Panel (c), which corresponds to the most emission signal, exhibits the largest gold region, where the estimated velocity lines up well with the true velocity despite starting with a mismatched velocity prior.\vspace{-15pt}}
    \label{fig:velocity_signal_ablation}
\end{figure}

We would like to verify that the velocity recovered by our method is similar to the true one that was used to simulate the ground-truth emissivity fields.
Panel (b) in \cref{fig:velocity_signal_ablation} shows the estimated velocity corresponding to the emissivity results shown in \cref{fig:emissivity}(a).
We plot the radial and azimuthal velocities as a function of radius.
Importantly, \pidef recovers radial and azimuthal velocities that more closely align with the true velocities than with the incorrect prior.

Since we lower the velocity regularization strength to essentially zero by the end of optimization, the recovered velocity is primarily fit to the measurements rather than the assumed velocity prior.
The recovered velocity is less accurate in regions where there is little emissivity.
This makes sense because there are no constraints on the velocity where there is little moving material.
\cref{fig:velocity_signal_ablation} shows how increasing the emissivity signal -- by either increasing the amount of gas present in the ground-truth or increasing the amount of measurement signal -- improves the velocity reconstruction accuracy.
We highlight a ``region of high emissivity density,'' defined as the region where the overall density of estimated emissivity (integrated along the entire time window) is above the $65$th percentile.
We find that the recovered velocity is more reliable in this region.

Outside of regions of high emissivity, the velocity recovery is fairly unconstrained.
We observe that radial velocity is harder to recover than azimuthal velocity and that velocity at small radius is difficult to recover.
As emission approaches the black hole, it is dimmed by the redshift factor $g^2$, which approaches zero on the event horizon. Additionally, infalling material is even more strongly redshifted as it moves away from the observer. As a result, emission from the infalling region close to the horizon makes only a very small contribution to the total flux density captured on the image plane, which likely explains why dynamics at small radius are difficult to reconstruct.


\begin{figure}
    \centering
    \includegraphics[width=\linewidth]{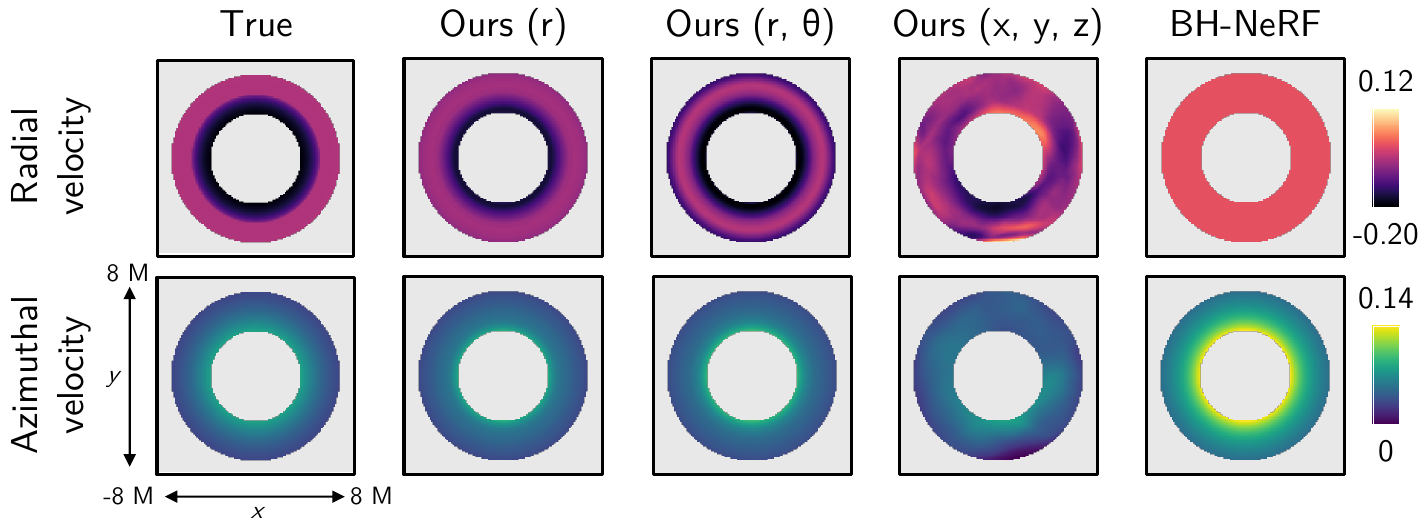}
    \caption{Axial symmetry is not an overly-restrictive constraint on the velocity network. Here we compare velocity reconstructions on the accretion disk for different DOFs in the velocity network. These velocity predictions correspond to the emissivity field in \cref{fig:emissivity}(a), using full-image measurements. These plots show the region of high emissivity density, defined to be greater than the $65$th percentile of emissivity densities, computed based on the $(r)$ reconstruction. The $(x,y,z)$ velocity contains more noise, which perhaps is higher where there is little reconstructed emissivity.\vspace{-15pt}}
    \label{fig:velocity_coordinate}
\end{figure}

\subsection{Parameterization of the velocity network}
The true velocity model is axially symmetric, and we can enforce this constraint on the velocity network by making it only depend on the radius $r$.
So far we have presented results using a velocity network that only depends on $r$.
In \cref{fig:velocity_coordinate}, we show what happens if we increase the number of degrees of freedom (DOFs) in the estimated velocity field.
We compare velocity networks that depend on $r$, $(r,\theta)$, and $(x,y,z)$ (while keeping the same true velocity that only depends on $r$).
We find reasonable velocity reconstructions for all three parameterizations when looking at the accretion disk in 2D, indicating that incorporating axial symmetry into the velocity network is not critical for our results.

\subsection{Towards real applications}
\label{subsec:realistic}
\begin{figure}
    \centering
    \includegraphics[width=0.95\linewidth]{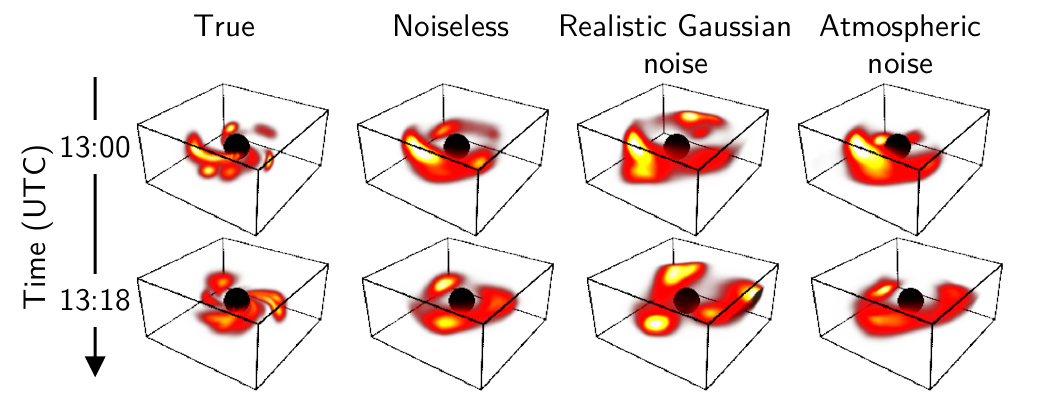}
    \caption{Emissivity reconstructions in challenging measurement settings. We simulated measurements assuming the ngEHT array. When assuming a realistic total flux of Sgr A*, the measured visibilities experience realistic Gaussian thermal noise. We also consider using closure phases (to overcome atmospheric noise) and amplitudes for data-fitting instead of complex visibilities. For this latter case, we assumed a large total flux that corresponds to effectively-noiseless measurements. Even so, using closure phases instead of visibility phases reduces reconstruction accuracy.}
    \label{fig:realistic}
\end{figure}

\paragraph{Realistic Gaussian noise}
We simulated EHT measurements of the same ground-truth emissivity as in \cref{fig:emissivity}(a) but with realistic Gaussian noise.
We rescaled the flux of the emissivity to have a mean total flux in the image plane of about $2.3$ Jy, the mean total flux of Sgr A* video reconstructions \citep{sgrapaperiii}, leading to approximately the same signal-to-noise ratio as expected for real EHT measurements of Sgr A*.
\cref{fig:realistic} shows a couple frames of the reconstruction.

\vspace{-10pt}
\paragraph{Atmospheric noise}
Phase errors caused by atmospheric turbulence can be difficult to remove through calibration.
\cref{fig:realistic} shows results of fitting to closure phases and amplitudes as measured by the ngEHT array rather than complex visibilities.
Closure phases are a nonlinear function of complex visibilities but circumvent station-based phase errors (more details are in the supplementary text).
In this experiment, we assumed noiseless EHT measurements since we only wanted to compare the performance of using closure phases and amplitudes versus complex visibilities.

\begin{figure}
    \centering
    \includegraphics[width=0.6\linewidth]{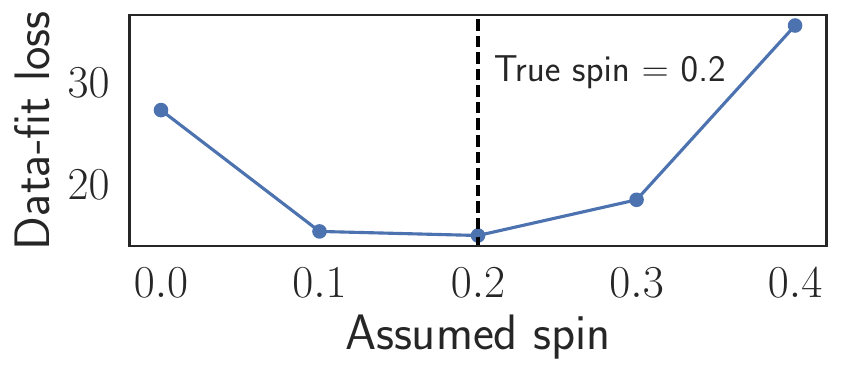}
    \caption{Data fit is sensitive to assumed spin. We compute data-fit loss as $\frac{1}{2\sigma^2}\left\lVert\mathbf{y}-\mathbf{A}\hat{\mathbf{x}}\right\rVert_2^2$, where $\hat{\mathbf{x}}$ is the estimated 4D emissivity field. For this proof-of-concept experiment, we assumed ngEHT measurements with realistic Gaussian noise with a correct velocity prior while varying the assumed spin (the ground-truth spin is $0.2$). This result demonstrates the possibility of inferring physics parameters via \pidef.\vspace{-10pt}}
    \label{fig:spin_sensitivity}
\end{figure}

\vspace{-10pt}
\paragraph{Spin sensitivity}
It may be possible to use our reconstruction approach to make scientific claims about the properties of the observed black hole.
We demonstrate this idea with a simple experiment showing that it may be possible to jointly infer the spin of the black hole along with the emissivity and velocity fields.
\cref{fig:spin_sensitivity} shows that the data-fit loss is sensitive to the assumed spin.
Assuming the correct spin of $0.2$ leads to the best data fit.
Although our optimization approach currently assumes a fixed spin and inclination angle, it is possible to jointly optimize these parameters.
The spin is especially important because it dictates the structure of spacetime and determines how scientists should study general relativity and quantum field theory near a black hole.

\section{Conclusion}
We have presented \pidef, a physics-informed approach for solving a highly ill-posed 4D tomography problem given sparse EHT measurements.
Our method recovers the 4D emissivity and 3D velocity fields of the gas near a black hole, a region of the universe that is ripe with potential physics discoveries.
Our experiments in simulation verify that \pidef provides superior reconstructions compared to the previous method, BH-NeRF. 
Furthermore, we are able to recover accurate dynamics where there is sufficient emissivity signal.
Our proof-of-concept experiments demonstrate the promise of \pidef to handle realistic noise, potentially positioning it as a practical tool for near-term science.
More broadly, our work opens the door to jointly inferring physical parameters, such as the spin of the black hole, and advances our ability to probe some of the most extreme environments in the universe.


\section*{Acknowledgments}
The authors would like to thank Abhishek Joshi for helpful discussions about GRMHD simulations. This work was supported by the National Science Foundation (NSF) under Cooperative Agreement PHY-2019786 (The NSF AI Institute for Artificial Intelligence and Fundamental Interactions), as well as NSF awards 1935980 and 2048237 and a Caltech Center for Sensing to Intelligence (S2I) award. BTF is supported by the NSF IAIFI and Tayebati Postdoctoral Fellowships and was partially funded by a Pritzker AI+Science Award. 

{
    \small
    \bibliographystyle{ieeenat_fullname}
    \bibliography{main}
}

\clearpage
\setcounter{page}{1}
\maketitlesupplementary

\appendix
\startcontents[appendices]
\printcontents[appendices]{}{1}{\setcounter{tocdepth}{2}}

\section{Velocity mismatch ablation}
\label{app:velocity_mismatch}
\begin{figure*}
    \centering
    \includegraphics[width=\linewidth]{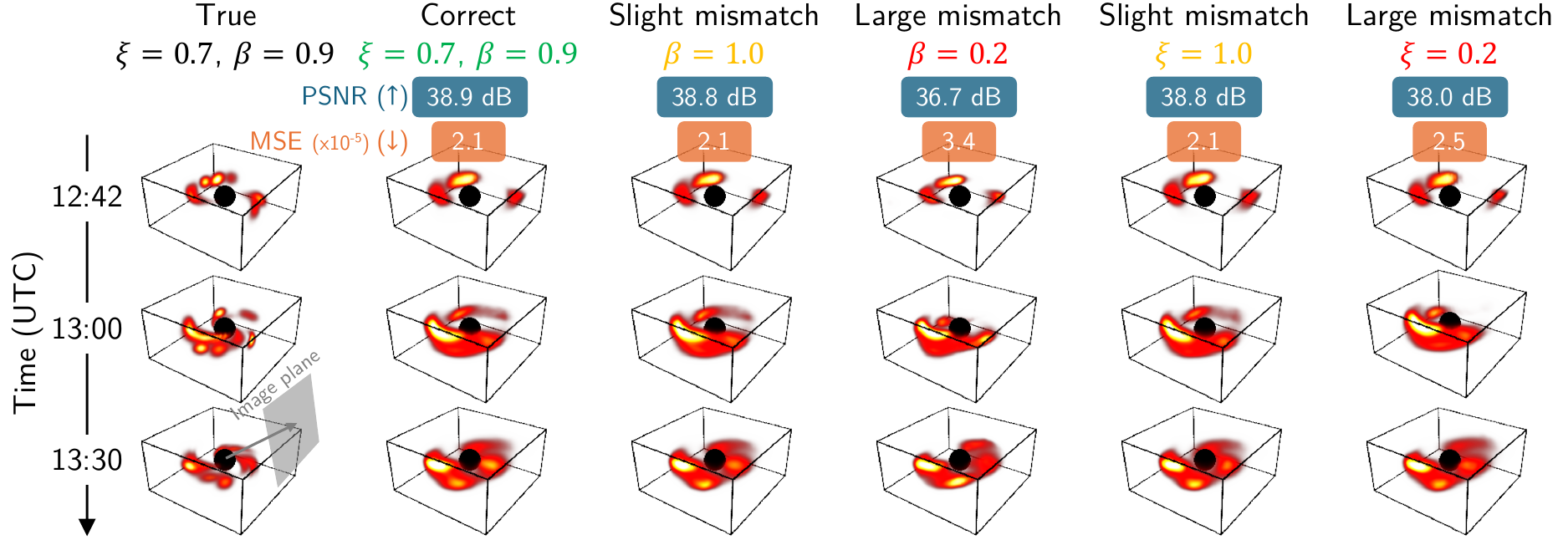}
    \caption{Ablation of the amount of mismatch in the velocity prior used for velocity regularization.}
    \label{fig:velocity_mismatch}
\end{figure*}
We performed an ablation study of the error in the assumed velocity model $v^i_\mathrm{prior}$.
\cref{fig:velocity_mismatch} shows emissivity reconstruction results when assuming different velocity priors with varying amounts of mismatch.
The ground-truth velocity follows the AART model with $\xi=0.7$ and $\beta=\beta_r=\beta_\phi=0.9$.
We ran optimizations for the following scenarios:
\begin{itemize}
    \item Correct velocity: $\xi_\mathrm{prior}=0.7$, $\beta_\mathrm{prior}=0.9$
    \item Slight mismatch in $\beta$: $\xi_\mathrm{prior}=0.7$, $\beta=1.0$
    \item Large mismatch in $\beta$: $\xi_\mathrm{prior}=0.7$, $\beta=0.2$ 
    \item Slight mismatch in $\xi$: $\xi_\mathrm{prior}=1.0$, $\beta=0.9$
    \item Large mismatch in $\xi$: $\xi_\mathrm{prior}=0.2$, $\beta=0.9$
\end{itemize}

A slight mismatch does not notably degrade reconstruction, and even a large mismatch still leads to a decent reconstruction.
These results indicate that our method is robust to errors in the assumed velocity model.
The main reason is that velocity regularization guides the emissivity and velocity reconstructions at the beginning of optimization, but as we gradually decrease the regularization strength, the reconstruction is mostly informed by the measurements.
\section{Implementation details}
\label{app:implementation}
Here we provide implementation details in addition to those provided in \cref{sec:implementation}.

\vspace{-10pt}
\paragraph{Simulating emissivity fields}
To create ground-truth emissivity fields, we introduced $N_\mathrm{flare}=20$ ``flares'' at evenly-spaced intervals from time $t=12.5$ to $t=13.5$ hours (UTC).
Each flare consisted of $1$, $2$, or $3$ (sampled uniformly) hotspots.
We randomly sampled hotspots with $r_{\min}=7$ and $r_{\max}=8$ and $\sigma_{\min}=0.5$ and $\sigma_{\max}=1.0$.
We propagated the hotspots according to the AART velocity model with $\xi=0.7$ and $\beta=\beta_r=\beta_\phi=0.9$.

\vspace{-10pt}
\paragraph{Simulating EHT measurements}
We used the \texttt{eht-imaging} library to simulate time-varying EHT measurements of the ground-truth emissivity fields.
\cref{tab:ehtparams} lists the parameter values used for simulating the EHT observation.
The observation time window of $12.5$ to $13.5$ hours (UTC) was chosen because $12.5$ to $14.2$ is considered a good observation time for Sgr A*.
We set the pixel size ($\mathrm{psize}$) parameter based on a field of view of $25$ M (about $125.345$ $\mu$as) and $100$ pixels in each direction in the image.

\vspace{-10pt}
\paragraph{Network architecture}
We used MLPs with $4$ layers with ReLU activations.
Each layer was $256$ units wide in the emissivity network and $128$ units wide in the velocity network.
We set the position-encoding degree as $L=3$ for the emissivity network and $L=1$ for the velocity network.
The emissivity 4D MLP and velocity MLP had 212,225 and 50,819 parameters, respectively.

\subsection{Optimization settings}
\paragraph{Hyperparameters} We set $\lambda_\mathrm{data}=1$ and $\Delta t=0.01$ for all measurement settings.
As mentioned in \cref{sec:implementation}, we gradually decreased the velocity regularization weight from $\lambda_\mathrm{init}$ to $\lambda_\mathrm{final}$.
When fitting to EHT measurements with negligible noise, we set $\lambda_\mathrm{dyn}=10^5$ and $\lambda_\mathrm{init}=10^6,\lambda_\mathrm{final}=10$.
When fitting to EHT measurements with realistic Gaussian noise, we set $\lambda_\mathrm{dyn}=10$ and $\lambda_\mathrm{init}=100,\lambda_\mathrm{final}=10^{-4}$.
When fitting to full images, we set $\lambda_\mathrm{dyn}=0.1$ and $\lambda_\mathrm{init}=1,\lambda_\mathrm{final}=10^{-6}$.

\vspace{-10pt}
\paragraph{Runtime}
On 1 NVIDIA A100 80GB GPU with simulated ngEHT data, it took 28 hours to run 100K optimization steps with a batch size of 6.
A ``batch'' refers to the number of time frames used to compute the data-fit loss, defined in \cref{eq:data_fit_loss}, which takes an expectation over the measurement time frames, and the dynamics loss, defined in \cref{eq:dynamics_loss}, which takes an expectation over the time interval $[0, T]$.

\vspace{-10pt}
\paragraph{ODE solver} To solve the integral in \cref{eq:integral}, we used the Tsitouras 5/4 solver \citep{tsitouras2011runge} implemented in Diffrax \citep{kidger2021on} with a constant step size of 0.001.

\begin{table}
\label{tab:ehtparams}
\centering
\begin{tabular}{ll}
\hline
\textbf{Parameter} & \textbf{Value} \\
\hline
$\mathrm{ra}$            & 17.761120 \\
$\mathrm{dec}$           & $-29.007797$ \\
$\mathrm{rf}$            & $230\times10^{9}$ \\
$\mathrm{bw}$            & $2\times10^{9}$ \\
$\mathrm{mjd}$           & 60775 (10/04/2025) \\
$\mathrm{source}$        & SgrA \\
$\mathrm{tstart}$        & 12.5 \\
$\mathrm{tstop}$         & 13.5 \\
$\mathrm{tint}$          & 102 \\
$\mathrm{tadv}$          & 102 \\
$\mathrm{tau}$           & 0.1 \\
$\mathrm{taup}$          & 0.1 \\
$\mathrm{polrep\_obs}$   & stokes \\
$\mathrm{elevmin}$       & 10.0 \\
$\mathrm{elevmax}$       & 85.0 \\
\hline
\end{tabular}
\caption{EHT measurement parameters.}
\end{table}
\section{EHT measurements}
\label{app:eht}
Here we provide details on very-long-baseline-interferometry (VLBI), which is the technique that the EHT uses to obtain Fourier measurements of the sky's image.
We denote the image by $I(x,y)$, where $(x,y)$ are 2D image coordinates.
The van Cittert-Zernike Theorem~\citep{van1934wahrscheinliche,zernike1938concept} states that the ideal visibility $v^*_{ij}$ measured by the baseline $\vb_{ij}$ between telescopes $i$ and $j$ is a single $(u,v)$ measurement on the complex 2D Fourier plane~\citep{thompson2017interferometry}:
\begin{align}
\label{eq:ideal_vis}
    v^*_{ij}:=\tilde{I}(u,v)=\int\int I(x,y)e^{-2\pi \text{i}(xu +yv)}\mathrm{d}x \mathrm{d}y.
\end{align}
An array of $N_s$ telescopes has 
$\binom{N_s}{2}$ independent baselines, each providing a visibility at each point in time.

In practice, there are multiple sources of noise in the measured visibilities.
Baseline-dependent \textbf{thermal noise} is modeled as a Gaussian random variable $\varepsilon_{ij}\sim\mathcal{N}(0,\sigma_{ij}^2)$, where $\sigma_{ij}$ is based on the system equivalent flux density (SEFD) of each telescope: $\sigma_{ij}\propto\sqrt{\text{SEFD}_i+\text{SEFD}_j}$.
The \textbf{station-dependent gain error} $g_i$ arises from each telescope $i$ using its own time-dependent $2\times 2$ Jones matrix \citep{hamaker1996understanding}. The \textbf{station-dependent phase error} $\phi_i$ arises from atmospheric turbulence that causes light to travel at different velocities toward each telescope~\citep{hinder1970observations,kolmogorov1997local,taylor1997spectrum}.
Other sources of corruption, including polarization leakage and bandpass errors, may introduce baseline-dependent errors, but they are slow-varying and assumed to be removable with \textit{a priori} calibration~\citep{chael2018interferometric}.
The measured visibility of baseline $\vb_{ij}$ can be written as
\begin{align}
\label{eq:meas_vis}
    v_{ij} = g_ig_j e^{\text{i}(\phi_i-\phi_j)}v^*_{ij}+\varepsilon_{ij}.
\end{align}

\subsection{Closure phases}
It is possible to deal with station-dependent errors by using \textit{closure quantities} -- specifically closure phases and closure amplitudes -- that are robust to such errors.
In \cref{subsec:realistic} we overcome atmospheric noise by using closure phases \citep{jennison1958phase}, which are formed by multiplying the three baselines within each triangle of telescopes $i,j,k$:
\begin{align}
    v_{ij}v_{jk}v_{ki}=&\left(g_ig_j e^{\text{i}(\phi_i-\phi_j)}v^*_{ij}+\varepsilon_{ij}\right)\notag \\
    &\left(g_jg_k e^{\text{i}(\phi_j-\phi_k)}v^*_{jk}+\varepsilon_{jk}\right) \notag \\
    &\left(g_kg_i e^{\text{i}(\phi_k-\phi_i)}v^*_{ki}+\varepsilon_{ki}\right) \\
    =&g_{ijk}^2 e^{\text{i}(\phi_i-\phi_j)}e^{\text{i}(\phi_j-\phi_k)}e^{\text{i}(\phi_k-\phi_i)}v^*_{ij}v^*_{jk}v^*_{ki}+\varepsilon_{ijk} \\
    =&g_{ijk}^2v^*_{ij}v^*_{jk}v^*_{ki}+\varepsilon_{ijk},
    \label{eq:bispectrum}
\end{align}
where $\varepsilon_{ijk}$ is a Gaussian random variable.
Although there are $\binom{N_s}{3}$ possible triplets in a telescope array, there are $\binom{N_s-1}{2}$ linearly independent closure phases.


\section{Details about black-hole emission physics}
\label{app:radiative_transfer}
Here we provide more detailed formulae for the quantities that we use to model fluid velocity and radiative transfer near black holes.
A mathematical object known as a \textit{spacetime metric} captures the geometry of the warped spacetime around a black hole.
It is represented as a \textit{metric tensor} $g_{\mu\nu}$, which allows us to define distances between points in spacetime, which are denoted by $x^\mu$.

\paragraph{Preliminaries}
In general relativity, we work with \textit{contravariant} (index-up) vectors $a^\mu$ and \textit{covariant} (index-down) vectors $a_\mu$. 
To lower indices, we multiply a contravariant vector by the spacetime metric:
\begin{equation}
    a_\mu=g_{\mu\nu}a^\nu.
\end{equation}
Conversely, to raise indices, we multiply a covariant vector by the inverse metric:
\begin{equation}
    b^\mu=g^{\mu\nu}b_\nu.
\end{equation}
Here $\mu,\nu\in(0,1,2,3)$ are indices.
We can compute dot products between pairs of covariant and contravariant vectors:
\begin{equation}
    a\cdot b = a^\mu b_\mu =  a_\mu b^\mu = \sum_{i=0}^3 a_ib^i.
\end{equation}

\subsection{Converting from the four-velocity}
Recall from \cref{sec:fluid_velocity} that we denote the three-velocity vector by
\begin{equation}
\label{eq:vi2}
    v^i=\left(v^r, v^\theta, v^\phi\right) = \left(\frac{\mathrm{d}r}{\mathrm{d}t},\frac{\mathrm{d}\theta}{\mathrm{d}t},\frac{\mathrm{d}\phi}{\mathrm{d}t}\right)
\end{equation}
and the four-velocity vector by
\begin{equation}
\label{eq:umu2}
u^\mu = \left(\frac{\mathrm{d}t}{\mathrm{d}\tau}, \frac{\mathrm{d}r}{\mathrm{d}\tau},\frac{\mathrm{d}\theta}{\mathrm{d}\tau},\frac{\mathrm{d}\phi}{\mathrm{d}\tau}\right) = u^t\left(1,v^r,v^\theta,v^\phi\right).
\end{equation}
The four-velocity $u^\mu$ can be directly derived from $v^i$ via Equation \ref{eq:umu2}, with
\begin{equation}
\label{eq:ut}
    u^t = \sqrt{\frac{-1}{g_{tt} + 2 g_{ti}v^i + g_{ij}v^iv^j}}
\end{equation}
to satisfy the normalization condition $u^\mu u_\mu =-1$.

\subsection{Converting from the normal observer frame}
As discussed in \cref{sec:normal_observer}, we estimate velocities in the normal observer frame to avoid numerical instabilities.
We denote the three-velocity in the normal observer frame by $\tilde{u}^i$.
The conversion from $\tilde{u}^i$ to $u^\mu$ is given by 
\begin{align}
    u^t &= \frac{\gamma}{\alpha}, \\
    u^i &= \tilde{u}^i-\left(\frac{\gamma}{\alpha}\right)\beta^i, \label{eq:normal_observer_conversion}
\end{align}
where $i$ is an index into $(r,\theta,\phi)$.
The lapse $\alpha$ and shift vector $\beta^i$ are defined as
\begin{align}
    \alpha &= \sqrt{\frac{1}{-g^{tt}}}, \\ 
    \beta^i &= -\frac{g^{ti}}{g^{tt}}.
\end{align}
The Lorentz factor $\gamma\geq1$ can be computed from the normal-observer $\tilde{u}^i$ as
\begin{equation}
    \gamma = \sqrt{1+g_{ij}\tilde{u}^i\tilde{u}^j}.
\end{equation}

The condition for the four-velocity to be physical is $\gamma > 1$.
In Boyer-Lindquist coordinates in the Kerr spacetime, we have that
\begin{equation}
\gamma^2 = 1 + \frac{\Sigma}{\Delta}\left(u^r\right)^2 + \Sigma \left(u^\theta\right)^2 + \frac{\Xi\sin^2\theta}{\Sigma}\left(u^\phi\right)^2.
\end{equation}
By inspection, we can see that $\gamma^2\geq1$ for all $u^i$.
There may be numerical instabilities when $\Delta\rightarrow0$, which happens at the horizon.

The conversion can be done in the other direction by solving for $\tilde{u}^i$ given $u^\mu$ in Equation \ref{eq:normal_observer_conversion}.
Recall that $u^\mu$ can be determined from $v^i$ by computing $u^t$ with Equation \ref{eq:ut}.

\subsection{AART velocity model}
Recall from \cref{sec:fluid_velocity} that we work with the AART velocity model \citet{cardenas-avendano_adaptive_2023}.
The model depends on a fixed spin $a$ and mass $M$, and it is axially symmetric, meaning in the equatorial plane it only depends on the radius $r$.
The AART model mixes two velocities: (1)
a sub-Keplerian velocity denoted by $u^\mu_\mathrm{subkep}$ and (2) an infall velocity denoted by $u^\mu_\mathrm{infall}$.
The velocity $u^\mu_\mathrm{subkep}$ is based on the Cunningham \citep{cunningham_effects_1975} model of Keplerian dynamics, which includes infall inside the innermost stable circular orbit (ISCO).
The velocity $u^\mu_\mathrm{infall}$ only represents infall due to geodesics, assuming a particle that starts with zero velocity at radius infinity.
Here we provide formulae for $u^\mu_\mathrm{subkep}$ and $u^\mu_\mathrm{infall}$, the derivations of which can be found in Appendix F of \citep{chael_black_2023}.
In the rest of this appendix, we will refer to the following common abbreviations in the Kerr metric:
\begin{align}
    \Delta &= r^2+a^2-2Mr, \\
    \Sigma &= r^2+a^2\cos^2\theta, \\
    \Xi &= \left(r^2+a^2\right)^2-a^2\Delta\sin^2\theta, \\
    \Omega &= \frac{2Mar}{(r^2+a^2)^2-a^2\Delta\sin^2\theta}.
\end{align}

\paragraph{Sub-Keplerian velocity model}
The Cunningham model \citep{cunningham_effects_1975} treats the velocity differently depending on whether it is outside or inside the ISCO.
The ISCO radius is given by
\begin{align}
    r_{\text{ISCO}} &= M\left[3 + Z_2 \pm \sqrt{(3-Z_1)(3+Z_1+2Z_2)}\right],
\end{align}
where
\begin{align}
    Z_1 &= 1 + (1-a^2/M^2)^{1/3}\left[(1+a/M)^{1/3}+(1-a/M)^{1/3}\right], \\
    Z_2 &= \left(3a^2/M^2 + Z_1^2\right)^{1/2}.
\end{align}
We define the following quantities:
\begin{align}
    \lambda_r&=\xi\cdot\frac{\text{sign}(a)\cdot s \cdot (r^2+a^2-2s\lvert a \rvert \sqrt{r})}{r^{3/2}-2\sqrt{r}+s\lvert a \rvert}, \\
    \gamma_r&=\sqrt{\frac{\Delta_r}{\Xi_r/r^2-4a\lambda_r/r-(1-2/r)\lambda_r^2}},
\end{align}
where $\xi\in(0,1]$ is a sub-Keplerianity parameter.
Here $s=\pm 1$ signifies whether the orbit is in prograde ($s=1$) or retrograde ($s=-1$).
We define $\lambda^*=\lambda_{r_\mathrm{ISCO}}$ and $\gamma^*=\gamma_{r_\mathrm{ISCO}}$.
The formulae for the four-velocity components depend on whether $r$ is outside or inside the ISCO, with the main difference coming from negative radial velocity inside the ISCO.
The components are computed as
\begin{align}
    u^t_\mathrm{subkep}(r) &=
    \begin{cases}
        \frac{\gamma_r}{\chi_r}, \quad &r\geq r_\mathrm{ISCO} \\
        \frac{\gamma^*}{\chi^*_r}, &r\leq r_\mathrm{ISCO}
    \end{cases}, \\
    u^r_\mathrm{subkep}(r) &=
    \begin{cases}
        0, & r\geq r_\mathrm{ISCO} \\
        -\frac{1}{r^2}\gamma^*\Delta_r\nu^*_r, & r < r_\mathrm{ISCO}
    \end{cases}, \\
    u^\theta_\mathrm{subkep}(r) &= 0 \\
    u^\phi_\mathrm{subkep}(r) &=
    \begin{cases}
        u^t_\mathrm{subkep}\Omega_r,& r\geq r_\mathrm{ISCO} \\
        u^t_\mathrm{subkep}\Omega^*_r, & r < r_\mathrm{ISCO}
    \end{cases},
\end{align}
where we use the subscript $r$ to denote a dependence on the radius $r$.
Here
\begin{align}
    H_r &= \frac{2r-a\lambda_r}{\Delta_r}, \\
    \chi_r &= \frac{1}{1+\frac{2}{r}(1+H_r)},
\end{align}
and
\begin{align}
    H^*_r &= \frac{2r-a\lambda^*}{\Delta_r}, \\
    \chi^*_r &= \frac{1}{1+\frac{2}{r}(1+H^*_r)}, \\
    \nu^*_r &= \frac{r}{\Delta_r}\sqrt{\frac{\Xi_r}{r^2}-\frac{4a\lambda^*}{r}-\left(1-\frac{2}{r}\right)(\lambda^*)^2-\frac{\Delta_r}{(\gamma^*)^2}}.
\end{align}
The azimuthal velocities are defined as
\begin{equation}
    \Omega_r=\frac{\chi_r}{r^2}(\lambda_r+aH_r)
\end{equation}
and
\begin{equation}
    \Omega^*_r=\frac{\chi^*_r}{r^2}(\lambda^*+aH^*_r).
\end{equation}

\paragraph{Infall velocity model}
The four-velocity components due to geodesic infall from infinity are given by
\begin{align}
    u^t_\mathrm{infall}(r) &= \frac{\Xi_r}{r^2\Delta_r}, \\
    u^r_\mathrm{infall}(r) &= -\frac{\sqrt{2r(r^2+a^2)}}{r^2}, \\
    u^\theta_\mathrm{infall}(r) &= 0, \\
    u^\phi_\mathrm{infall}(r) &= \frac{2a}{r\Delta_r}.
\end{align}

\paragraph{Putting everything together}
We define $\Omega_\mathrm{subkep}=\frac{u^\phi_\mathrm{subkep}(r)}{u^t_\mathrm{subkep}(r)}$ and $\Omega_\mathrm{infall}=\frac{u^\phi_\mathrm{infall}(r)}{u^t_\mathrm{infall}(r)}$.
The four-velocity components of the general velocity model are given by
\begin{align}
    u^t(r) &= \frac{1+\frac{r^2(u^r)^2}{\Delta_r}}{1-(r^2+a^2)\Omega^2-\frac{2}{r}(1-a\Omega)^2}, \\
    u^r(r) &= \beta_ru^r_\mathrm{subkep}+(1-\beta_r)u^r_\mathrm{infall}, \\
    u^\theta(r) &= 0, \\
    u^\phi(r) &= u^t(r)\cdot\left(\beta_\phi\Omega_\mathrm{subkep}+(1-\beta_\phi)\Omega_\mathrm{infall}\right).
\end{align}
The parameters $\beta_\phi,\beta_r\in[0,1]$ determine the amount of influence of the sub-Keplerian model in the azimuthal and radial velocities, respectively.

\subsection{Redshift}
As mentioned in \cref{sec:image_formation}, there is a Doppler effect on the intensity of light that reaches the observer.
The redshift factor depends on the velocity field.
Assuming a photon energy of $E=1$, the location-dependent redshift factor $g$ is computed as
\begin{align}
\label{eq:redshift}
    g = \frac{E}{-k_\mu u^\mu} = \frac{1}{-k_\mu u^\mu},
\end{align}
where $k_\mu$ is the photon momentum vector.
It is related to the derivative of the photon position $x^\mu(\tau)$ with respect to the Mino time $\tau$:
\begin{equation}
    \frac{\mathrm{d}x^\mu}{\mathrm{d}\tau}=\frac{\Sigma}{E}k^\mu=\Sigma k^\mu = \Sigma g^{\mu\nu}k_\nu
\end{equation}
assuming $E=1$.
For a given spin and mass of the black hole, the photon momentum at every point in spacetime is fixed.


\end{document}